\documentclass[Afour,sageh,times]{sagej}

\usepackage{moreverb,url}
\usepackage{breqn}
\usepackage{amsmath}
\usepackage{amsmath}
\usepackage{graphicx}
\usepackage[colorinlistoftodos]{todonotes}
\usepackage{float}
\usepackage{subfigure}
\usepackage[english]{babel}
\usepackage[utf8x]{inputenc}
\usepackage[T1]{fontenc}
\usepackage{nomencl}

\newcommand\BibTeX{{\rmfamily B\kern-.05em \textsc{i\kern-.025em b}\kern-.08em
T\kern-.1667em\lower.7ex\hbox{E}\kern-.125emX}}

\makenomenclature

\usepackage[numbers]{natbib}

\begin{document}

\runninghead{Davoudi and Morris}

\title{Self-Noise modelling and acoustic scaling of an axial fan configured with rotating controlled diffusion blade}

\author{Behdad Davoudi\affilnum{1} and Scott C. Morris\affilnum{2}}

\affiliation{\affilnum{1}Michigan State University, East Lansing, MI, USA -- currently at Aerospace Engineering Department, University of Michigan, Ann Arbor, MI, 48108 USA\\
\affilnum{2}Department of Aerospace and Mechanical Engineering, Notre Dame, IN, 46556 USA
}

\email{davoudi@umich.edu, s.morris@nd.edu}

\begin{abstract}
A semi-empirical acoustic model for self-noise was adapted to predict the sound radiated from an axial fan featuring rotating controlled diffusion blades (RCDB). Experimental data for wake velocity, mass flow rate across the fan, and fan rotational speed were obtained. These experimental data along with typical characteristics of turbulent boundary layers were used to predict the noise of the axial fan. Hot-wire wake measurements were made in the near region downstream of the fan plane. 
The fan noise was measured upstream of the fan. The experimentally obtained self-noise was then compared to the predictions made by the semi-empirical acoustic model. Goody's and Rozenberg's models for surface pressure spectra were used in the semi-empirical acoustic model. Rozenberg's model offered a more accurate prediction in the final fan noise spectra. The predictions were in reasonable agreement with experimental data, and the model was found to be a useful tool to reasonably estimate acoustic emissions of a fan with limited information about the velocity field in the fan wake. Different operating conditions and blade configurations were examined. For a given dimensionless operating condition, the self-noise was obtained for different rotational speeds, and the effect of the fan speed on the propagated noise was evaluated. The acoustic scaling function was experimentally obtained as a function of normalized frequency and dimensionless operating condition and it was found to be quite frequency dependent despite it is often assumed as a constant value.
\end{abstract}


\nomenclature{$[\hspace{0.1cm}]_{rms}$}{Root mean squared of [ ]}
\nomenclature{$B$}{Number of blades}
\nomenclature{$c$}{Blade Chord length}
\nomenclature{$c_0$}{Sound speed}
\nomenclature{$f$}{Frequency}
\nomenclature{$H$}{Shape factor}
\nomenclature{$L_3$}{Blade span}
\nomenclature{$\dot{m}$}{Mass flow rate}
\nomenclature{$M_c$}{Convective Mach number}
\nomenclature{$P_{atm}$}{Atmospheric pressure}
\nomenclature{$r$}{Radial direction}
\nomenclature{$R_T$}{Inner-layer to outer layer time scale}
\nomenclature{$\bar{U}$}{Spatial and temporal mean axial velocity}
\nomenclature{$v'$}{Velocity fluctuations}
\nomenclature{$V_{tip}$}{Blade tip velocity}
\nomenclature{$\delta$}{Boundary layer thickness}
\nomenclature{$\delta^*$}{Displacement Thickness}
\nomenclature{$\delta^{**}$}{Boundary layer momentum thickness}
\nomenclature{$\Delta P$}{Pressure differential across the fan }
\nomenclature{$\Lambda_3$}{Span-wise integral length scale }
\nomenclature{$\phi$}{Flow coefficient}
\nomenclature{$\Phi_{pp}$}{Wall pressure auto spectral density}
\nomenclature{$\Phi_{p,rad}$}{Radiated acoustic auto spectral density}
\nomenclature{$\tau_{wall}$}{Wall shear stress}
\nomenclature{$\psi$}{Pressure rise coefficient}
\nomenclature{$\omega$}{Angular frequency}
\nomenclature{$\Omega$}{Rotational speed}
\nomenclature{$\beta_c$}{Equilibrium parameter}

\maketitle
\printnomenclature

\section{Introduction}

Axial fans are widely used for relatively high volume flow rate and low pressure rise applications. cooling fans and ventilation systems are common application areas. Current and pending regulations on noise emissions make noise measurements and modeling for axial fans important topical areas for original equipment manufacturers and suppliers. This paper presents measurements and modeling for a three and a nine blade axial fan. The cross section of the fan blades was designed with a Controlled Diffusion (CD) airfoil. The CD airfoil controls the boundary layer growth on the suction side that is of importance in axial fan design. Surface pressure fluctuations of the CD airfoil have been studied in \cite{wang2009prediction,moreau2005effect}. Aeroacoustic characteristics of the CD airfoil were investigated in \cite{moreau2005effect,christophe2011uncertainty,sanjose2011direct} focusing mainly on trailing edge noise which is identified as the primary high-frequency noise source in an airfoil. 

The characteristics of the blade wakes and blade surface pressure for the three blade axial fan was investigated by \cite{neal2010effects}. For the same configuration, Cawood \cite{cawood2012surface} specifically investigated the surface pressure fluctuations on the fan blades. Davoudi \cite{davoudi2014aeroacoustic} conducted acoustic measurements and quantified the velocity fluctuations downstream of the fan for the three and nine blade configurations. Barrent \cite{barrent2015controlled} obtained comprehensive fan performance measurements including wake and blade surface pressure for multiple blade configurations. A study of the nine-blade version of this axial fan, including both the acoustics and fluid mechanics characteristics, is reported in \cite{davoudi2016self}.\\ 

The noise generated by an axial fan in the absence of inflow turbulence and distortion is termed fan self-noise. Self-noise represents distinct noise sources such as Turbulent Boundary Layer Trailing Edge (TBL-TE) noise, separation noise, tip leakage noise, vortex shedding noise, etc. It is difficult to quantitatively determine the amplitude of each of type of acoustic source since there are usually a number of contributors in each experiment. TBL-TE noise is often considered to be an important source of the overall noise generation. Semi-empirical acoustic models often utilize the surface pressure spectra upstream of the TE as an input to predict the self-noise; see Amiet \cite{amiet1976noise} and Blake \cite{blake1986mechanics}. 

An increase in pressure rise across an axial fan is generally associated with increase in noise generation \cite{longhouse1977vortex,stephens2011measurements}. Rotor sound spectra from a variety of axial flow machines (with different rotor sizes and tip speeds) have been compared by Wright \cite{wright1976acoustic}. His study showed that a common characteristic can describe various generated rotor noises. That is, an inverted parabolic curve in log-log coordinates can be a reasonable approximation to sound spectra. Wake measurements using stationary and rotating hot-wire probes as well as acoustic measurements were made for a small axial fan by Quinlan and Bent \cite{quinlan1998high}. They indicated that the tip flow contributes significantly to overall broad band noise for frequencies above 1 KHz.\\   

Longhouse \cite{longhouse1977vortex} considered the effect of rotational speed increase on the generated sound, and found that SPL is proportional to $v^n_{tip} r^2_{tip} $ where he reported the exponent $n$ to be approximately $3$ in the absence of upstream disturbances. A value of $5$ was obtained for lower flow rates. Mellin \cite{mellin1975selection} demonstrated a value of $5$ based on an analytical prediction. Stephens and Morris \cite{stephens2011measurements} obtained the exponent value as a function of frequency and indicated that the scaling factor is not constant for all frequencies. In the present study, the scaling factor as a function of frequency and dimensionless operating condition was obtained for the nine-bade RCDB axial fan.\\

An empirical self-noise prediction code was developed by Brooks \textit{et al.} \cite{brooks1989airfoil} based on the data base of airfoil noise measurements from Brooks and Marcolini \cite{brooks1985scaling}. Glegg and Jochault \cite{l1998broadband} used the noise prediction code to predict the self-noise of a ducted fan by use of an acoustic analogy with the Green's function. Rotor solidity, duct mode coupling and rotating sources were also taken into consideration. Their sound spectra predictions featured mass flow rate, blade twists and other parameters. Gliebe \cite{gliebe2002fan} extended a rotor TE noise model suggested by Mugridge and Morfey \cite{mugridge1972sources} based on his review of broadband mechanisms in aircraft engine fans. He indicates that a "noise floor" can be set based upon TBL-TE noise. \\

Acoustic and wake measurements are reported for both three and nine blade configurations in this paper. An acoustic model adapted from Stephen and Morris \cite{stephens2011measurements} was used to predict the RCDB emitted noise. The experimental setup and the fans’ specifications are described in next section. Then, the fans’ performance curves and velocity fluctuations in the plane in close proximity of blades TE are presented. The wake velocity results as well as the blade-relative flow velocity (derived from the mass flow rate and rotational speed) are used as inputs to the acoustic semi-empirical model. The comparison between the noise predictions (by the semi-empirical acoustic model) and the measured spectra  and also acoustic scaling factor for the nine blade fan follow in later sections.

\section{Experimental Setup and Details}
\subsection{Experiment}

Accurate measurements of flow rate (through the fan plane), rotational speed, wake velocity and emitted noise (in the region upstream of the fan) can be obtained using the Axial Fan Research and Development (AFRD) Facility of the Turbulent Shear Flows Laboratory at Michigan State University, see~\cite{morris2001moment} for experimental details.

\begin{figure}[!htb]
\centering
\includegraphics[scale=.6]{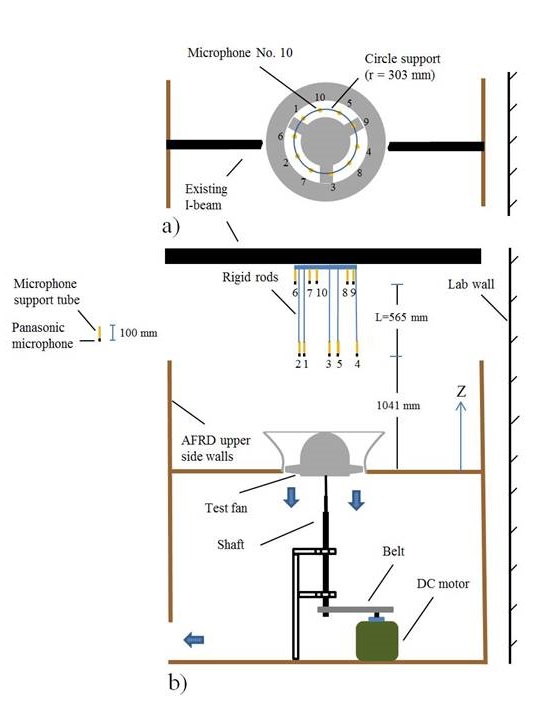}
\caption{The Axial Fan Research and Development (AFRD). a) Top view. b) Side view. }
\label{fig:AFRD}
\end{figure}

An overall schematic of the facility is shown in Fig.~\ref{fig:AFRD}. The airflow enters from the laboratory atmosphere to the upper receiver by passing through the fan plane.
A pressure tap embedded in the ceiling of the AFRD facility, was used to measure the differential pressure rise across the fan. The fan rotation rate was set by a $15$ horsepower DC motor. A once-per-revolution sensor was used measure the rotational speed. The clearance between the blade tips and the circular shroud was $4 mm$. A traverse mechanism positioned a single hot-wire probe in close proximity ($2-3 mm$) of the blade TE. The traverse system could accurately place the anemometer probe at different heights and span-wise locations. \\

It is noted that Davoudi \textit{et al} \cite{davoudi2016aeroacoustic} showed that the aerodynamic noise of a nine-blade axial fan (the same fan studied herein) dominates the motor and other facility noise. They operated the fan at $1000$ \textit{RPM} with zero blades installed on the hub, and compared the corresponding propagated noise with that of the full nine-blade fan assembly.

It is noted that the environment in which the acoustic measurements were made was not anechoic. A microphone installed in the fan upstream will receive the target sound (from the blades), coherent reflections from the surrounding walls, and also the ambient noise. This issue was be mitigated by measuring the acoustic field using multiple microphone locations simultaneously. This allows the use of post-processing techniques to reduce the effects of unwanted noise sources. 

Ten calibrated microphones were placed in two circular arrays upstream of the fan. The upper and the lower arrays contain 5 microphones that are evenly spaced every $72$ degrees with a stagger angle of $36$ degrees between them. The arrays were centered (above the blades’ mid-span) and placed at two different elevations (1041 mm and 1606 mm) above the fan plane; see Fig.~\ref{fig:AFRD}. Data were sampled at a rate of $70 KHz$ for $95$ seconds. Davoudi \cite{davoudi2016aeroacoustic} developed and demonstrated beamforming processing methods that was shown to provide an accurate measurement of the auto-spectral density of the acoustic pressure at a location upstream of the fan. Details of the microphones calibration and the employed beamforming technique to attenuate the extraneous noise are provided by Davoudi \textit{et al.}~\cite{davoudi2016aeroacoustic}.

\subsection{RCDB Fans Description}

\begin{figure}[!htb]
\centering
\includegraphics[scale=.25]{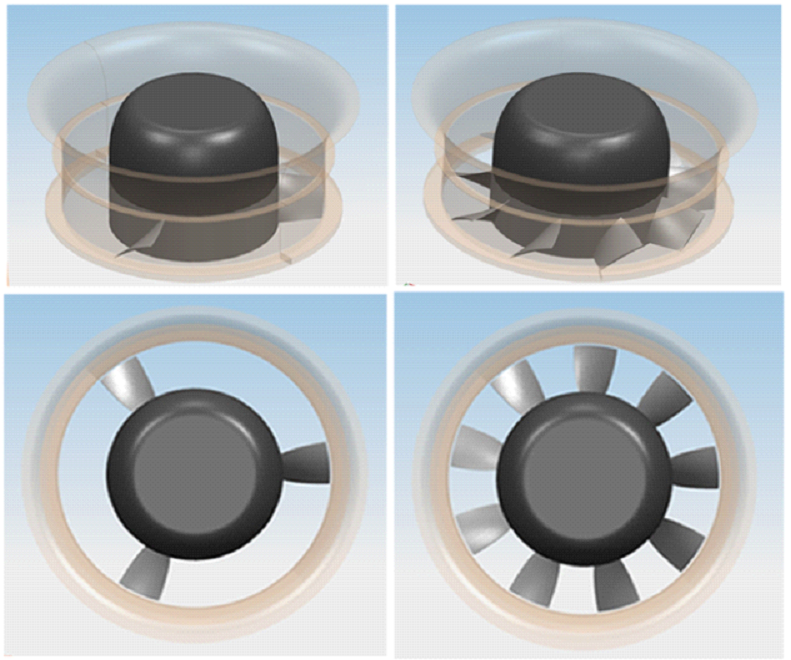}
\caption{CAD model images of the three (left) and nine (right) bladed fans.}
\label{fig:3_vs_9}
\end{figure}

 The axial fan blades have a specific Controlled Diffusion (CD) airfoil profile; the chord length ($c$) is $133.9 mm$, the thickness-to-chord ratio is $4\%$, and the camber angle is $12$\textdegree. The fan blades are referred to as Rotating Controlled Diffusion Blade (RCDB). Publications that have investigated the same CD profile were cited in the introduction. Schematics of the three and nine blade fans are depicted in Fig.~\ref{fig:3_vs_9}. RCDBs were made from identically cambered airfoils, and they are twisted along the span. See [10] for the twist specifications. The two blade configurations for the fans represent two distinct blade-to-blade interaction effects: i) minimal and ii) interacting. The fan hub radius was $r_{root}=0.240 m$ , and the tip radius was $r_{tip}=0.366 m$. A nominally uniform axial inlet velocity ($\bar{U}$) is expected given the contoured inner and outer shrouds.
 
\begin{figure}[!htb]
 \centering
 \subfigure[three blade fan selected: $\phi=0.36,0.26$]{
  \includegraphics[width=0.4\textwidth]{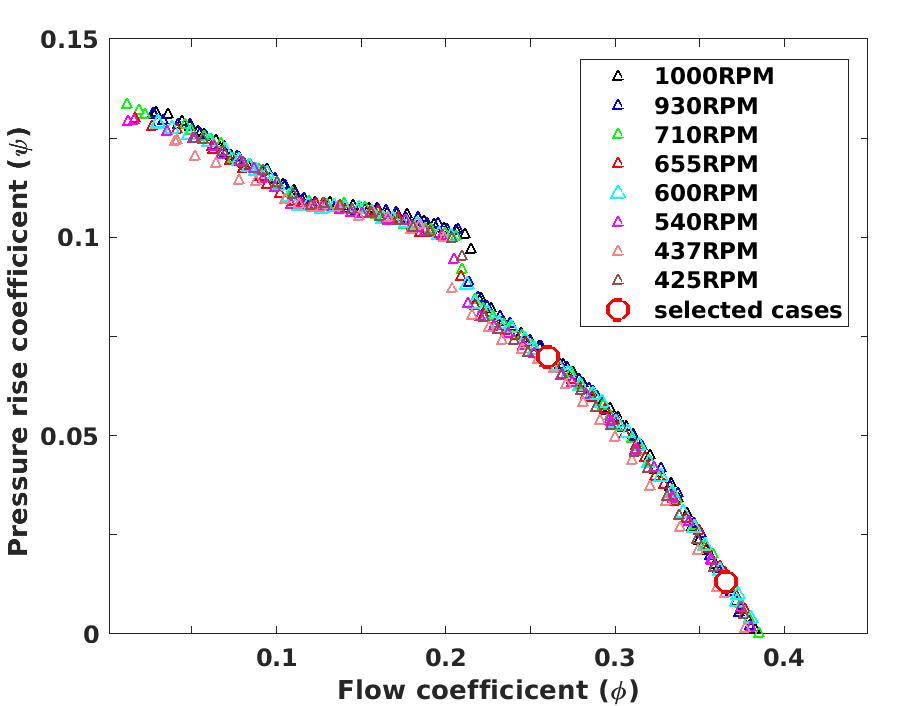}
 }
 \centering
 \subfigure[nine blade fan selected: $\phi=0.49,0.36,0.31,0.29$]{
  \includegraphics[width=0.4\textwidth]{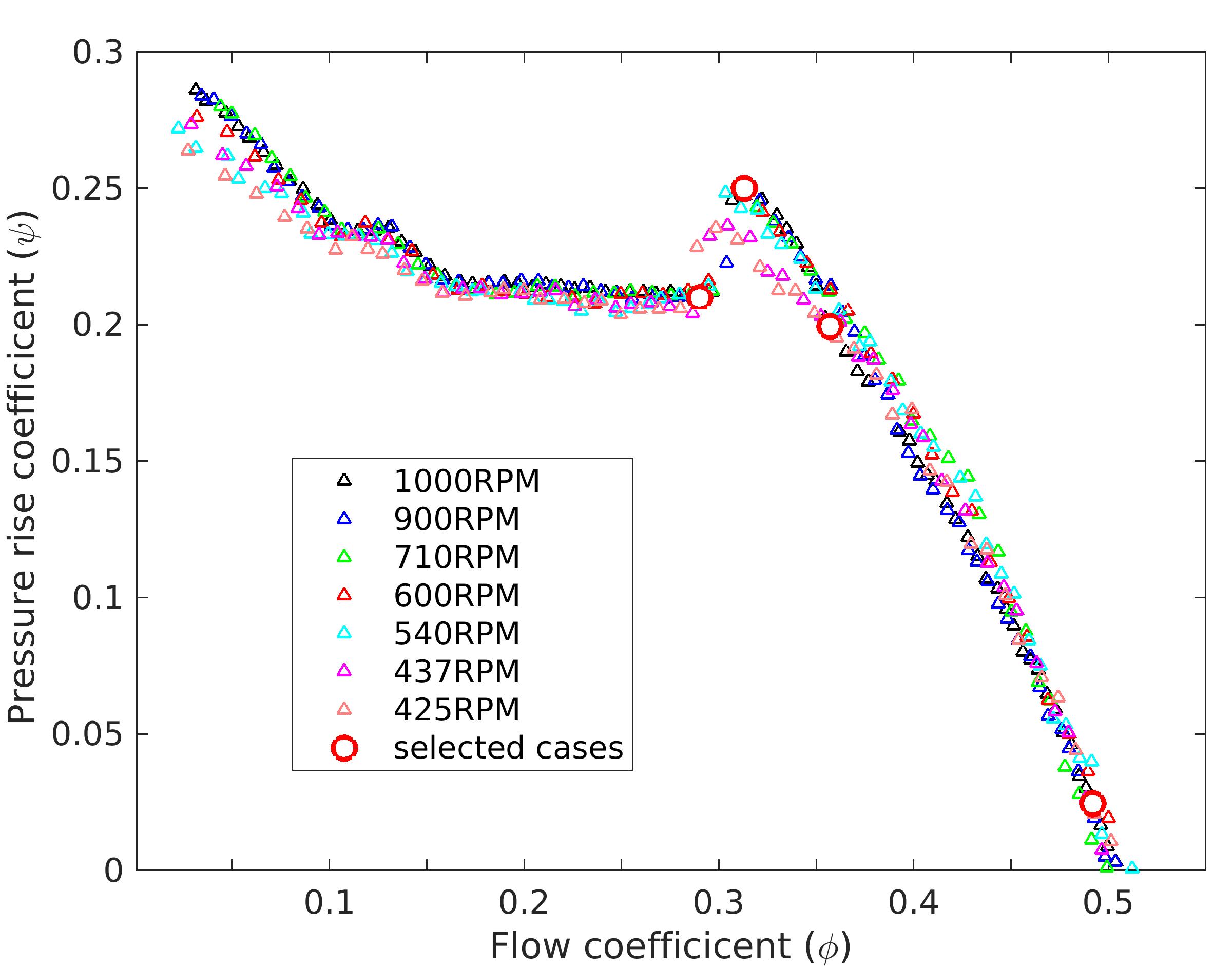}
 }
\caption{Non-dimensional performance curves}
\label{fig:per}
\end{figure}

The dimensionless characteristic curves for the fan is defined by the pressure rise coefficient ($\psi=2\Delta P/\rho v^2_{tip} $) and flow coefficient ($\phi=\bar{U}/v_{tip}$). Results for the nine and three blade fans are shown in Fig. \ref{fig:per} (reproduced from \cite{davoudi2014aeroacoustic}).  $\Delta P$ was measured as the differential between the static pressure in the downstream volume and the upstream total pressure. $\bar{U}$ is the spatially and temporally averaged axial velocity of the inflow ($\dot{m}/\rho A$), and  $v_{tip}$ is the blade tip velocity. $A$ is the fan plane annular area: $\pi [r^2_s-r^2_h]/4$, where $r_s=0.370 m$ and $r_h=0.240 m$ are the shroud and hub radii respectively.

 The three blade characteristics for 8 different rotational speeds are shown in Figure 3a. In dimensionless form all of the measurements collapse well to a single curve, indicating that the integral properties of the flow are almost insensitive to Reynolds number. The dimensionless flow rate through the fan was found to be approximately $\phi=0.38$ when the pressure rise was close to zero. The operating point with $\phi=0.36$ is marked on the figure, and was considered for more detailed study in the following sections. The 3-bladed fan increases pressure gradually as the flow rate was reduced. The condition $\phi=0.26$ is also marked on the characteristic and considered further below, as this represents a typical application design point for this type of fan. At lower flow rates a ``jump'' in pressure was observed at $\phi=0.21$. This was assumed to be related to a transition on the blades from a predominantly axial flow condition to a more radial exit flow field. 
 
 The nine blade fan characteristics are shown in Figure 3b. $\phi=0.49$ represents the flow rate observed an close to zero pressure rise. Several selected cases are noted on the figure, including $\phi=0.36$ which represents a typical system design. A discontinuity in pressure rise of the nine blade fan characteristic curve was observed at $\phi=0.31$ where a sudden decrease in the fan's pressure rise occurred.  This behavior was related to blade-stall (large scale flow separation) on the suction surface of the blade. This behavior is discussed in the next section using wake velocity data.

\section{Self-noise Modeling}

One objective of this paper is to apply a simplified modeling approach for the radiated sound generated by the fan. The approach will follow the methods originally introduced by Blake's \cite{blake1986mechanics} and Stephens and Morris \cite{stephens2011measurements}. The method assumed that the noise sources could be approximated by finite spanwise section, and then integrated in order to approximate the total source amplitude. The sources were approximated by the trailing edge noise model introduced by Blake \cite{blake1986mechanics}. Stephens and Morris \cite{stephens2011measurements} estimated the radiated noise of a fan where the fan blades were acoustically compact in the frequency range of interest ($c<\frac{c_0}{f}$). The same method was used in this paper; however, RCDB fan blades are not acoustically compact for the frequency range of interest, and some adjustments were made to estimate the non-compact range of interest that will be discussed in details.

\begin{figure}[H]
\centering
\includegraphics[scale=.35]{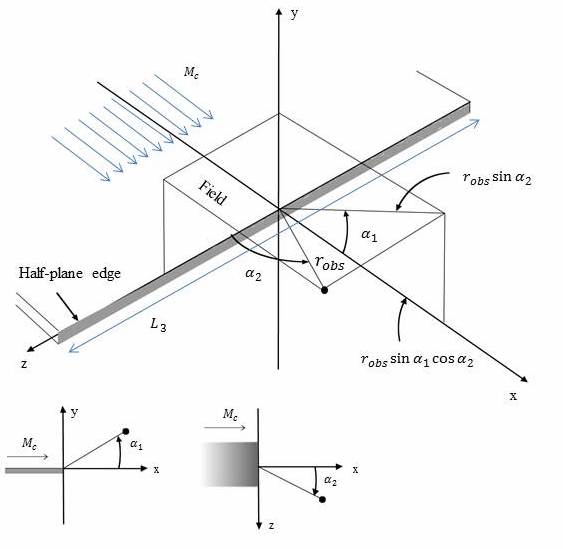}
\caption{Geometry of a dipole source near a rigid half-plane. (This is an idealization of the TE sound for non-compact radiation.)}
\label{fig:goody}
\end{figure}

Blake \cite{blake1986mechanics} developed a self-noise prediction model for the flow passing over one side of a semi-infinite half plane (i. e., a TE) as shown in Fig.~\ref{fig:goody}. His model accounts for turbulent boundary layer – Trailing edge (TBL-TE) noise which is mainly represented by a dipole-like sound source. The unsteady lift is associated with the unsteady surface pressure in a boundary layer. The model provides the radiated pressure auto-spectra ($\Phi_{p_{rad}}$) of a convecting flow passing over the TE for an observer located at the distance ($r_{obs}$) from the TE. The predicted spectra is function of unsteady surface pressure spectra ($\Phi_{pp}$), plane span ($L_3$), observer location with respect to flow direction and TE, and other turbulence quantities. His model follows as ($f<\frac{c_0}{c}$):

\begin{equation}
\label{eqn:goodyedge}
[\Phi_{p_{rad}}]_{edge}=
\Phi_{pp}(f)\frac{1}{4\pi^2}\frac{L_3\Lambda_3(f)}{|r_{obs}|^2}M_c sin^2 \frac{\alpha_1}{2} |sin \alpha_2| 
\end{equation}

\noindent where, $M_c$ is the convective Mach number ($M_c=U_c/c_0$). $c_0$ is the sound velocity. $\Lambda_3(f)$ is the span-wise integral length scale of the unsteady wall pressure. Blake \cite{blake1986mechanics} suggested $\Lambda_3(f)=U_c/2\pi f\gamma_3$ as it is difficult to measure. He also suggested that $U_c/U_{\infty}=0.7$ and $\gamma_3=0.8$. With these assumptions, one can obtain $\Lambda_3 \approx U_{\infty}/2\pi f$. The $U_{\infty}$ can be approximated by blade relative velocity that can be derived from mass flow rate and rotational speed measurements. The angles $\alpha_1$ and $\alpha_2$ account for the observer position with respect to flow and TE directions. As shown in Fig. ~\ref{fig:goody}, the former is defined between the stream-wise direction and the latter is defined between the span-wise direction and $r_{obs}$. It is appropriate to neglect the directivity effects in this study given that the self-noise is sought far enough above the fan blades and $\alpha_1$ and $\alpha_2$ are close to 90 degrees.
 
Eq. \ref{eqn:goodyedge} was developed for a semi-infinite half-plane edge. Thus, Eq.\ref{eqn:goodyedge} is valid if the chord length is larger than the acoustic wave length ($c_0/f$) which follows for frequency values greater than $c_0/c$ (acoustically non-compact). Using the strip theory, Stephens and Morris \cite{stephens2011measurements} integrated Eq. \ref{eqn:goodyedge} and obtained an equation for the self-noise of their acoustically compact fan blade in the frequency range of interest. Blake \cite{blake1986mechanics} also presented a relation that can provide the acoustic radiation from a compact airfoil based on the non-compact spectra ($[\Phi_{p_{rad}}]_{edge}$). If one neglects the directivity effects, there results:

\begin{equation}
\label{eqn:edgetocompcat}
\frac{[\Phi_{p_{rad}}]_{edge}}{[\Phi_{p_{rad}}]_{compact}}\cong\frac{8}{\pi(\frac{\omega c}{c_0})}  
\end{equation}

\noindent Accounting for compactness of an airfoil in the frequency range of interest, Eq.\ref{eqn:goodyedge} and Eq.\ref{eqn:edgetocompcat} follow as:

\begin{equation}
\label{eqn:ncompact}
[\Phi_{p_{rad}}]_{non/comp}=\Phi_{pp}(f)\frac{1}{4\pi^2}\frac{L_3\Lambda_3(f)}{|r_{obs}|^2}\frac{M_c}{2} \quad f>\frac{c_0}{c} 
\end{equation}

\begin{equation}
\label{eqn:compact}
[\Phi_{p_{rad}}]_{compa}=\Phi_{pp}(f)\frac{fc}{16c_0}\frac{L_3\Lambda_3(f)}{|r_{obs}|^2}\frac{M_c}{2} \quad f<\frac{c_0}{c}  
\end{equation}

\noindent The $c_0/C$ ratio is about 2.5 KHz for the RCDB ($c=133.9 mm$). Thus, the RCDB can be considered to be a compact airfoil below 2.5 KHz. The half-plane edge noise principle can be used for frequency values above this limit. 

In Eqs. \ref{eqn:ncompact} and \ref{eqn:compact}, the auto-spectral density of the surface pressure fluctuations can be obtained by time resolved surface pressure measurements. Alternatively, these values can be approximated using  semi-empirical models (i.e., \cite{goody2004empirical} ). Flow measurements to obtain boundary layer characteristics near the TE should be made to obtain the required inputs to Goody's model. The wake velocity measurements in this study were used to estimate the required inputs to Goody's model that will be discussed in the next chapter.

As indicated before, Blake’s model was defined for a stationary airfoil exposed to a uniform free stream. This can be used for a rotating airfoil using a ``strip-theory'' approach.  Specifically, the rotating blade can be divided into small spanwise segments with a width of $dr$.  The net noise source is then recovered by an summation of the contributions from each strip. The surface pressure spectra (proposed by Goody) are different for each radial location. Thus, Blake’s equation can be re-written for RCDB fan as:
\begin{equation}
\label{eqn:int_ncom}
[\Phi_{p_{rad}}]_{non/comp}=\frac{B}{4\pi^2|r_{obs}|^2}\int_{r_{hub}}^{r_{tip}}\left(\phi_{pp-ss}+\phi_{pp-ps}\right)
\end{equation}
\begin{equation*}
\Delta_3 (f,r) M_c(r)\ dr \quad f>\frac{c_0}{c}\qquad\qquad\qquad
\end{equation*}

\begin{equation}
\label{eqn:int_com}
[\Phi_{p_{rad}}]_{comp}=\frac{Bfc}{16c_0|r_{obs}|^2}\int_{r_{hub}}^{r_{tip}}\left( \phi_{pp-ss}+\phi_{pp-ps}\right)
\end{equation}
\begin{equation*}
\Delta_3 (f,r) M_c(r)\ dr \quad f<\frac{c_0}{c}\qquad\qquad\qquad
\end{equation*}\\

\noindent where, $\Delta_3(f,r)=W_\infty (r)/2\pi f$, $M_c(r)$ is the convective Mach number ($W_{\infty}(r)/c_0$) and $\Phi_{pp-ss} (f,r)$, $\Phi_{pp-ps} (f,r)$ are the suction side, pressure side surface pressure spectra. $B$ is also the number of blades.

Goody \cite{goody2004empirical} developed an empirical model for surface pressure spectra beneath a turbulent boundary layer with zero pressure gradient. His model is a result of various curve fitting procedures for several experimental data sets presenting a large range of Reynolds numbers, $1.4 \times 10^4 < Re_{\theta} < 2.34 \times 10^4$. Goody’s model follows as:

\begin{equation}
\label{eqn:goody}
 \frac{\phi_{PP} (\omega) U_e }{\tau^2_w}
=
\frac{3(\frac{\omega \delta}{U_e})^2}
{\left [ (\frac{\omega \delta}{U_e})^{0.75}+0.5\right ]^{3.7}
+\left [ R_T^{-0.57} (\frac{\omega \delta}{U_e})\right ]^7
}   
\end{equation}

\noindent in which, $\delta$ is the boundary layer thickness, $U_e$ is the velocity at the edge of the boundary layer, $\omega$ is the angular frequency, and $\tau_w$ is the wall stress. $R_T$ is the ratio of the timescales of the outer to inner boundary layer and is defined as:

\begin{equation}
\label{eqn:rt}
   R_T=\frac{\delta/U_e}{\nu/u_{\tau}^2} 
\end{equation}

where $u_{\tau}$ is the friction velocity. It is noted that time the scale ratio also contains the effect of Reynolds number. Goody [26] proposed $R_T=0.11(U_e\delta^{**}/\nu)$ where $\delta^{**}$ is the boundary layer momentum thickness.

Wake measurements were obtained in the fan downstream as will be discussed in details in the next section. The purpose of the measurements is to provide basic understanding of the flow field in order to interpret the acoustic measurements. Additionally, it will be assumed that the wake flow field (close to the TE of a blade) can be used to obtain estimates of the upstream boundary layer characteristics. Hence, the Goody model with wake values used as input parameters can be utilized to provide an approximate description of the surface pressure spectra near the TE of the RCDB. The required $U_e$ can be approximated by the blade relative velocity ($w_{\infty}=\sqrt{\bar{U}^2+(r\Omega)^2}$). Wall shear stress can be approximated with the maximum of the RMS of velocity fluctuations squared in the wake region of a blade \cite{stephens2011measurements} as $\tau_w=\rho[v^{'2}_{rms}]_{max}$. Similarly, the wake thickness can be used as an approximation to the boundary layer thickness for the model. This length scale can be obtained at each radial location as the circumferential distance between the points where the RMS of velocity fluctuations corresponded to half of the peak value. Davoudi \cite{davoudi2014aeroacoustic} used two separate approximations for the suction and pressure side boundary layers. That is, the suction side boundary layer is defined as the circumferential distance between points where the RMS of velocity fluctuations was equal to half of the peak value. Similarly, the pressure side boundary layer is defined using the pressure side of the wake pattern. Note that the TE thickness is about $1-1.5 mm$.

A few approximations must be made in order to obtain the required parameters to utilize Goody’s model. Spalart and Watmuff \cite{spalart1993experimental} reported shape factors ($H=\delta^*/\delta^{**}$, where is the boundary layer displacement thickness) in the range $1.5<H<1.65$ for low Reynolds number boundary layer measurements in the presence of an adverse pressure gradient. Stephens and Morris \cite{stephens2011measurements} assumed $H=1.45$ and $\Delta=\delta/\delta^*=8$ as the general boundary layer characteristic of an axial ducted fan. The same values were selected for the present study, and surface pressure spectra were obtained for the RCDB pressure and suction sides.

Rozenberg et al \cite{rozenberg2012wall} proposed a semi-empirical model based on Goody’s model in which Adverse Pressure Gradient (APG) effects were considered. Goody’s model takes only into account Reynolds number effects on the pressure spectra. Rozenberg examined his model versus experimental data for 6 different cases (3 pipe flows and 3 airfoils) in the presence of adverse pressure gradient. His model was found to be in a better agreement with the experiments especially for mid and low frequencies comparing to Goody’s model. That is, Goody’s model underestimated the spectrum in low and mid frequency in the presence of APG. The Rozenberg’s model collapses with Goody’s model in the lack of APG. Rozenberg’s model follows in Eq. \ref{eqn:rozenberg}:

\begin{equation}
\label{eqn:rozenberg}
\small
 \frac{\phi_{PP} (\omega) U_e }{\tau^2_w}
=
\frac{
\left[ 2.82 \Delta^2 (6.13 \Delta^{-0.75}+F_1)^{A_1}\right] \left[ 4.2 (\frac{\Pi}{\Delta})+1 \right](\frac{\omega \delta^*}{U_e})^2
}
{
\left [ (4.7 \frac{\omega \delta^*}{U_e})^{0.75}+F_1\right ]^{A_1}
+\left [8.8 R_T^{-0.57} (\frac{\omega \delta^*}{U_e})\right ]^{A_2}
}   
\end{equation}

where, $A_1=3.7+1.5 \beta_c$, $A_2=min(3,19/\sqrt{R_T})$, $ F_1=4.76 (\frac{1.4}{\Delta})^{0.75} [0.375A_1-1]$.

$\beta_c=\left(\delta^{**}/\tau_w\right) (dp/dx)$ is the equilibrium parameter for self-similarity determination defined by  Clauser \cite{francis1954turbulent}. $\Pi$ is the wake strength parameter. Durbin et al \cite{durbin2011statistical} has proposed the following empirical formula based on the Clauser’s parameter:
$\Pi=0.8 (\beta_c+0.5)^{3/4}$                                                            
Rozenberg’s model will be used to estimate pressure spectra in the RCDB suction side on which APG exists.

\begin{figure}[H]
\centering
\includegraphics[scale=.75]{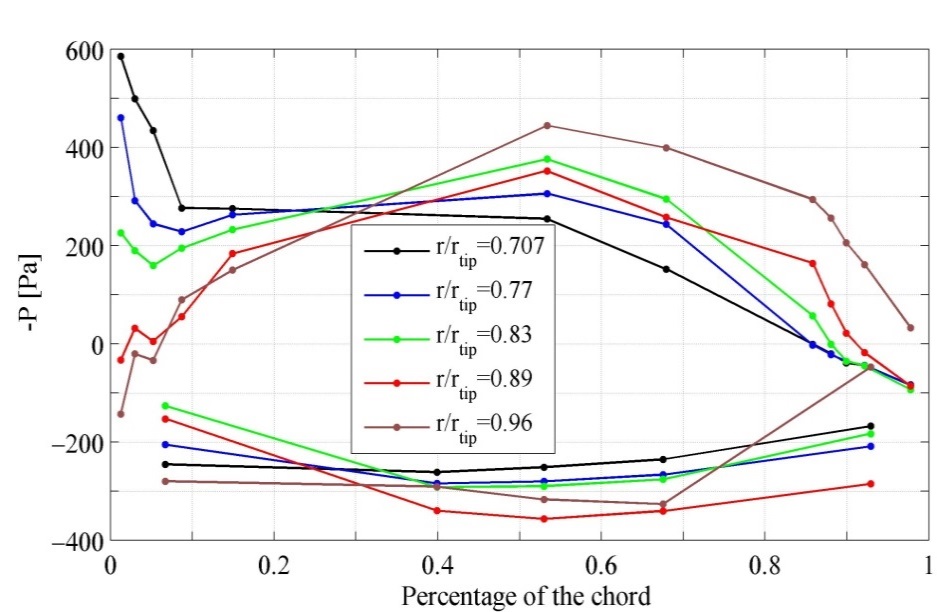}
\caption{pressure distribution on RCDB (9-bladed fan)}
\label{fig:pressure}
\end{figure} 
Barrent \cite{barrent2015controlled} has obtained surface pressure data for 9-blade fan at 450 RPM. The pressure values were scaled to estimate surface pressure at 1000 rpm. It is noted that the flow geometry is the same for a given inflow coefficient for different RPM conditions. To be brief, surface pressure values at 1000 RPM only for $\phi=0.49$ is provided in Fig.~\ref{fig:pressure} for different span-wise locations. Pressure gradient (dp/dx) was obtained for different span-wise location from Surface the pressure data.
\section{Wake Velocity Results}
Velocity measurements were obtained using single-sensor hot-wires at an axial plane that was 2mm downstream of the rotor trailing edge. The data are used here to study the flow field characteristics and to approximate the quantitative parameters needed for the acoustic model. 

 The sensors respond to the component of the air speed that is perpendicular to the hot-wire filament. Hence, it was necessary to orient the probe such that the hot-wire was normal to the mean flow angle of the air leaving the fan exit plane. This flow angle was found by first using a small tuft in order to document the mean flow angles. 

\begin{figure}[!htb]
 \centering
 \subfigure[$\phi=0.49$]{\label{fig:9case1}
  \includegraphics[width=0.47\textwidth]{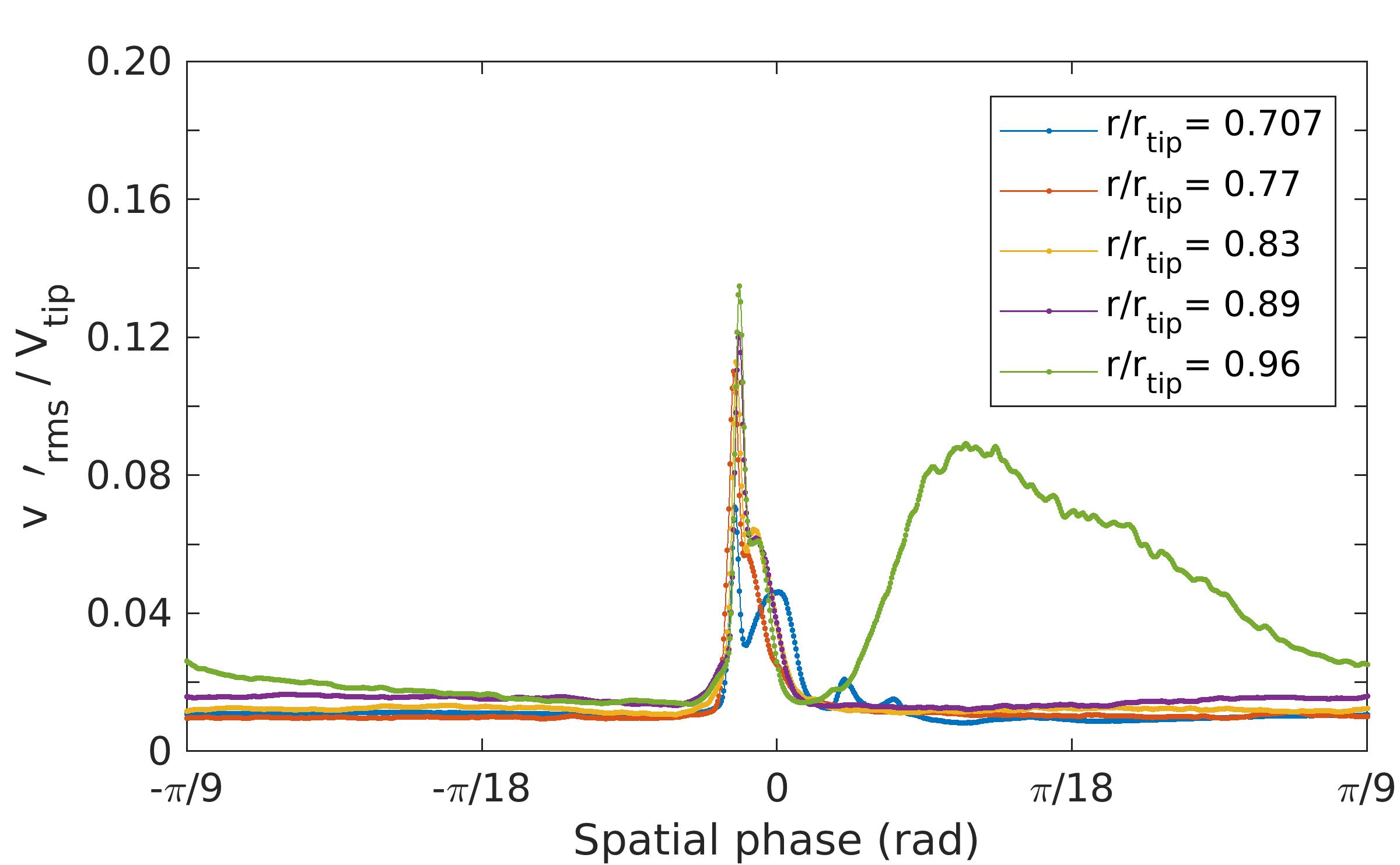}
 }
 \centering
 \subfigure[$\phi=0.36$]{\label{fig:9case2}
  \includegraphics[width=0.47\textwidth]{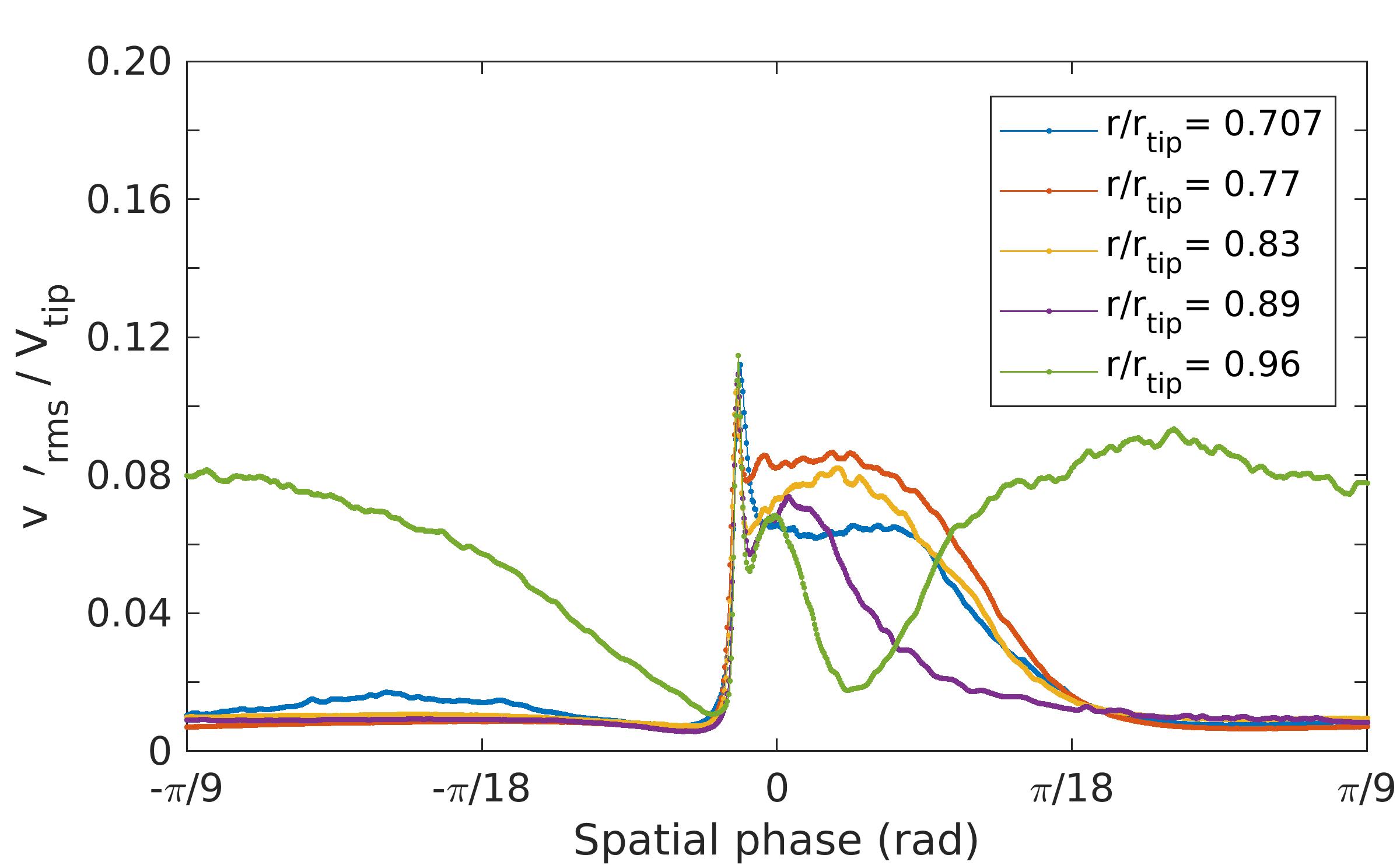}
 }
 \centering
 \subfigure[$\phi=0.31$]{\label{fig:9case3}
  \includegraphics[width=0.47\textwidth]{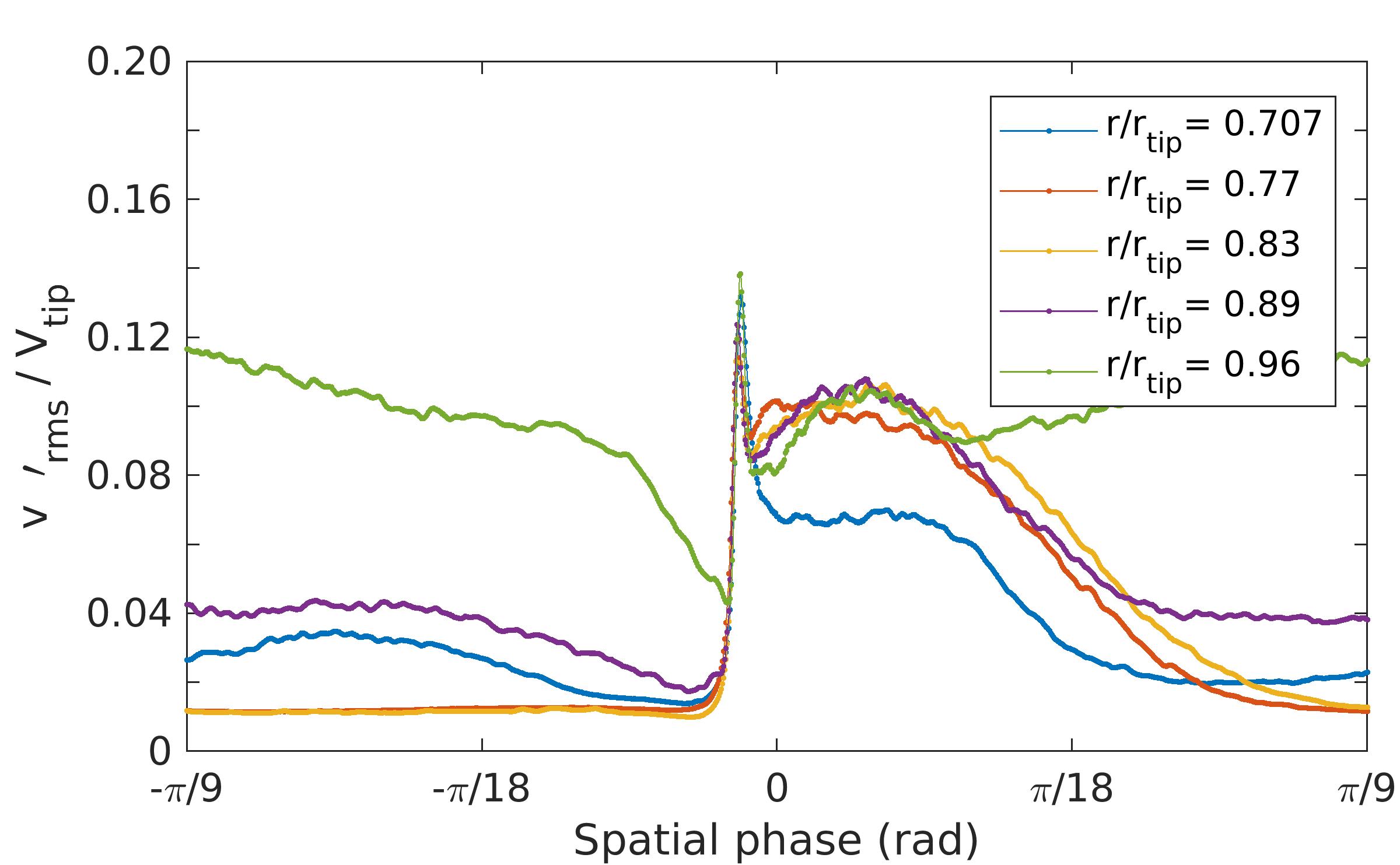}
 }
  \centering
 \subfigure[$\phi=0.29$]{\label{fig:9case4}
  \includegraphics[width=0.47\textwidth]{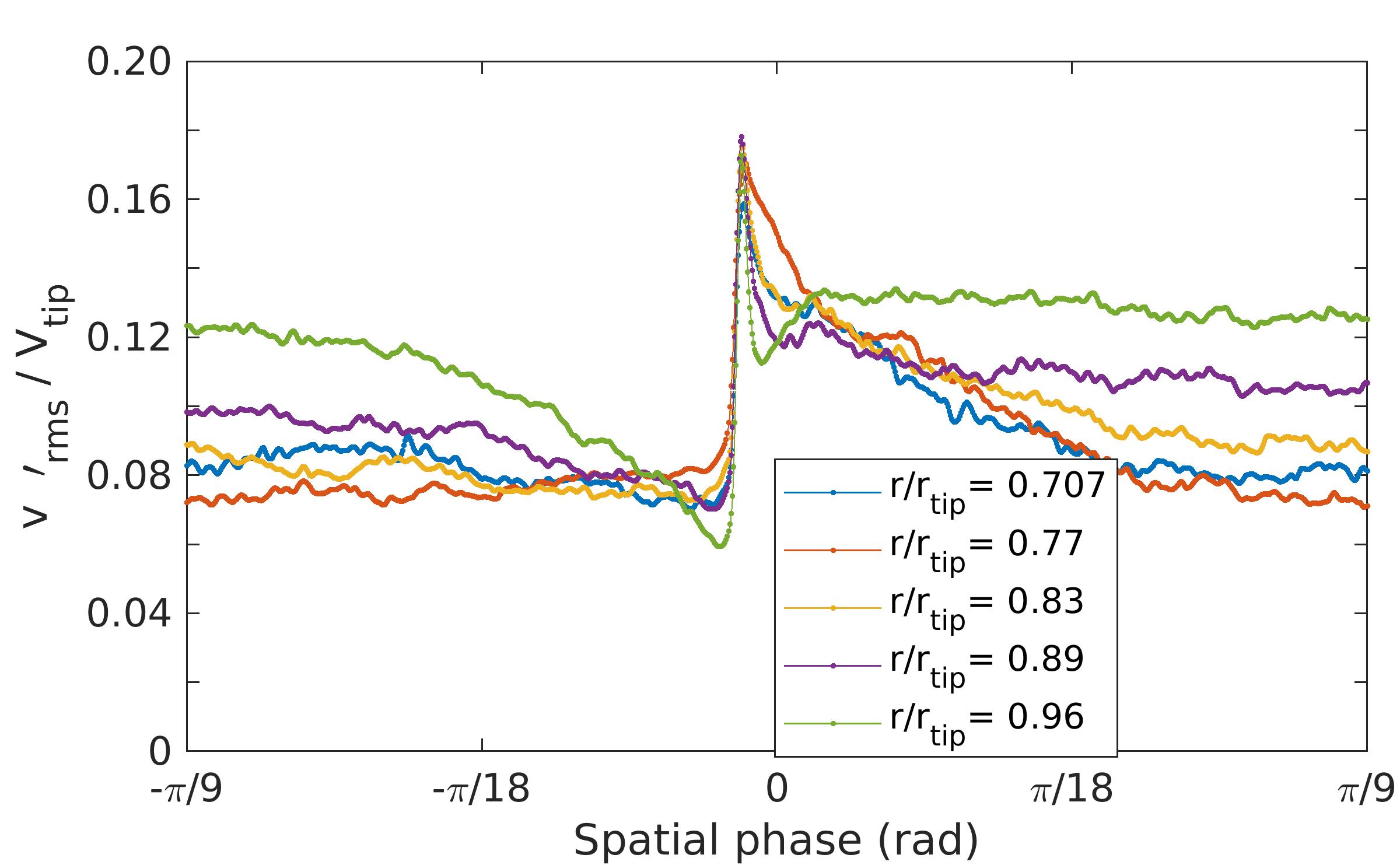}
 }
\caption{Velocity fluctuations level in the fan downstream - nine blade fan}
\label{fig:9cases}
\end{figure}

To obtain the velocity fluctuation level distributions in the wake region at $1000 RPM$, the time-series velocity data were phase-locked to the rotor, and RMS values of the difference between velocity data and phase-averaged mean velocity were computed at each rotor phase. Five different span-wise locations (starting from hub to tip) in the downstream region of the fan were selected for the wake measurements. ($r/r_{tip}= 0.707,0.77,0.83,0.89, 0.96$). The RMS of the velocity fluctuations for the nine blade fan, made non-dimensional by $V_{tip}$ are provided in Fig~\ref{fig:9cases} for different flow coefficients.

The wake velocity fluctuations for the nine bladed fan at a $\phi=0.49$ is shown in Fig.~\ref{fig:9case1}. The range of the abscissa corresponds to the pitch-wise distance between the blades (2$\pi$/9). The normalized RMS values were less than $0.02$ between the wakes (for all of the radial locations except for $r/r_{tip} = 0.96$). The wake can be observed as the narrow region of high RMS values near the middle of the figure. Two local maxima can be observed in several of the radial locations. The very sharp maxima is a result of the thin shear layer created from the pressure-side boundary layer. The second maximum (particularly at lower $r/r_tip$ values, is indicative of the thicker suction-side boundary layer. The elevated level at the outer radius ($r/r_{tip} = 0.96$) in between the blade wakes was a result of the tip clearance flow.
Similar results for $\phi=0.36$ are shown in Fig.~\ref{fig:9case2}. The narrow maxima corresponding to the shear flow generated by the pressure side boundary layer was found to be similar to the $\phi=0.49$ case. The second maxima corresponding to the suction side flow indicates a much larger sheared region with increase unsteadiness levels. The values in the pitch-wise region between the wakes was low (less than 0.02), with the exception of the $r/r_{tip} = 0.96$. This location illustrates that the entire flow passage between the blades was influenced by the tip flow turbulence.


Results for $\phi=0.31$ are plotted in Fig.~\ref{fig:9case3}. This condition represents a relatively high pressure rise, at a higher flow rate than the sudden decrease in pressure rise as noted on the fan characteristic curve. The narrow local maximum appears very similar to the previous two conditions. However, the region of unsteadiness on the suction side of the blades is significantly larger as might be expected for this high-pressure condition. 

The results for post-stall case ($\phi=0.29$) is shown in Fig.~\ref{fig:9case4}. The sharp local maxima at the blade pressure side remains evident at this condition, but the magnitude of the unsteadiness has increased compared with the higher flow rate cases. The unsteadiness can be observed to be elevated significantly across the remainder of the passage, with values in excess of 0.07 for all radial locations. This suggests large-scale separation on the suction side of the blades that affects the entire blade passage. A wake thickness cannot be defined for $\phi=0.29$ since the disturbed flow has occupied the entire passage between the blades.\\

\begin{figure}[!htb]
 \centering
 \subfigure[$\phi=0.36$]{\label{fig:3case1}
  \includegraphics[width=0.47\textwidth]{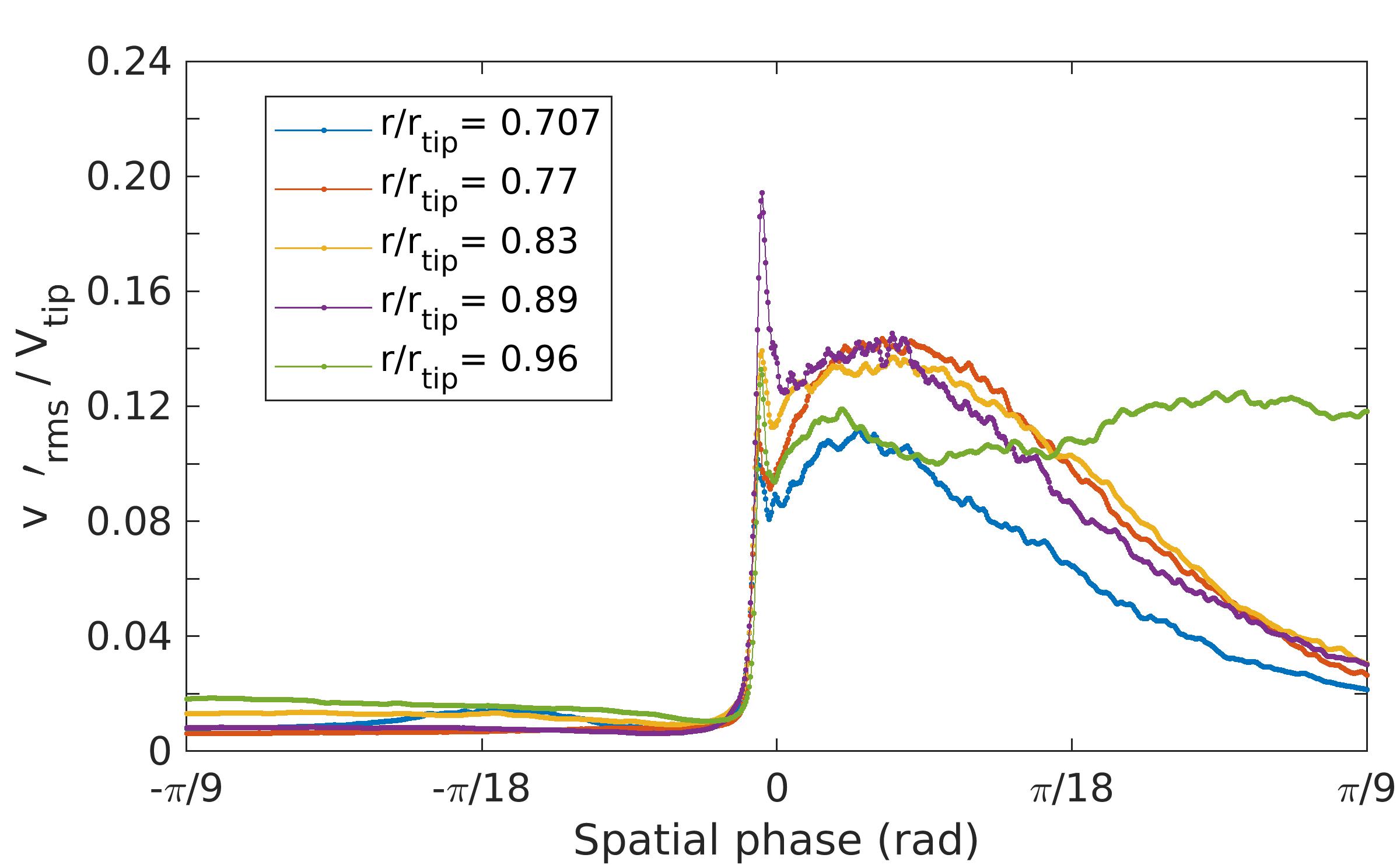}
 }
 \centering
 \subfigure[$\phi=0.26$]{\label{fig:3case2}
  \includegraphics[width=0.47\textwidth]{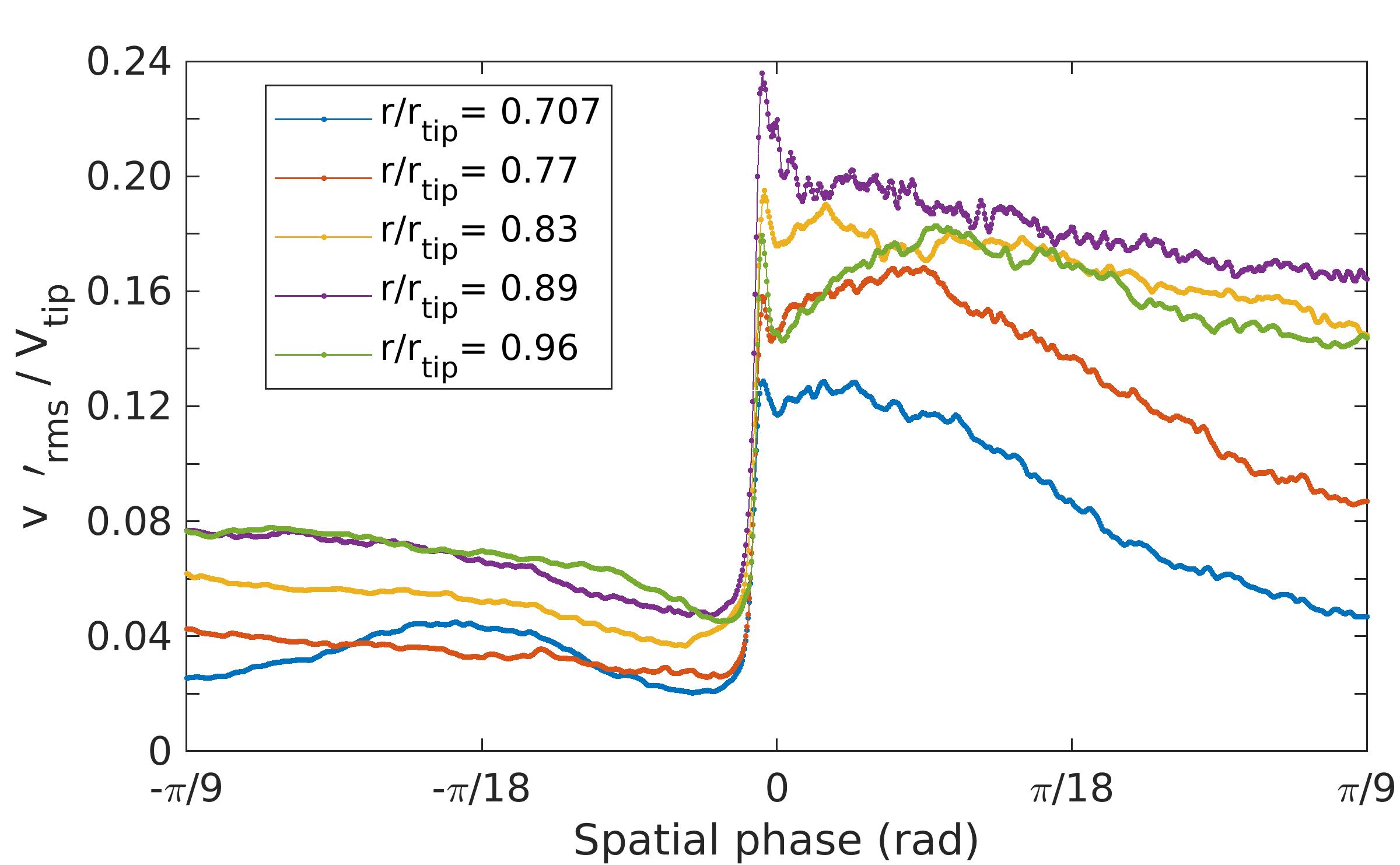}
 }
\caption{Velocity fluctuations level distribution in the fan downstream - three blade fan}
\label{fig:3cases}
\end{figure}

The velocity RMS values for the 3-bladed fan for $\phi=0.36$ and $\phi=0.26$ are shown in Figures \ref{fig:3case1} and \ref{fig:3case2}, respectively. The extent of the abscissa is the same as for the the 9 bladed case, but now represents one third of the blade pitchwise extent. At the higher flow rate, the results are similar to the 9-bladed fan, in that a narrow maxima is observed at all radial locations from the pressure-side shear flow. Similarly, the suction side flow results in a second, broader, local maximum. At the lower mass flow rate, increased unsteadiness values can be observed over a large extent on the suction side of the blades indicating that the airfoils are likely fully separated.  
Note that, a very thin boundary layer thickness on the pressure side of the blade is suggested by the data for all of the operating conditions.
\section{Acoustic Results and Discussion}
\begin{figure}[H]
\centering
\includegraphics[scale=.25]{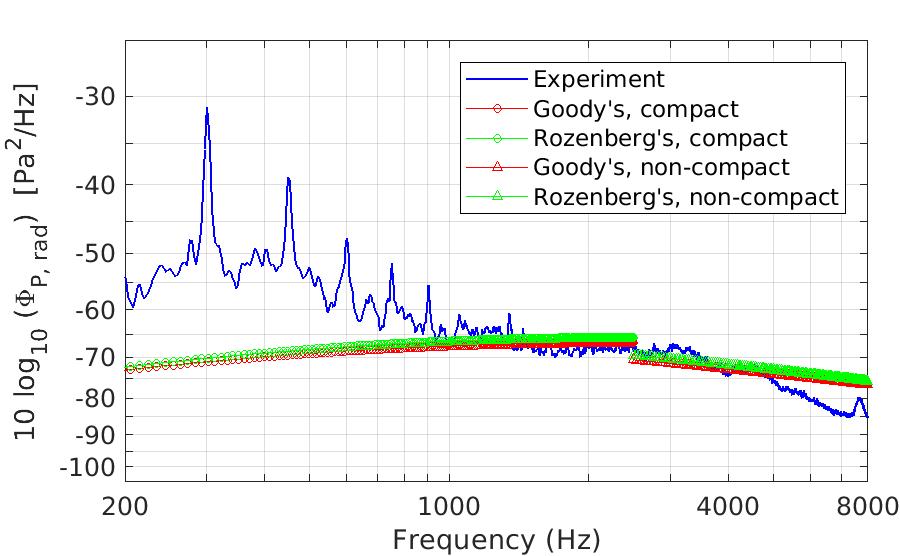}
\caption{Self noise predictions at $\Phi=0.49$  – nine blade fan}
\label{fig:self91}
\end{figure}

\begin{figure}[H]
\centering
\includegraphics[scale=.25]{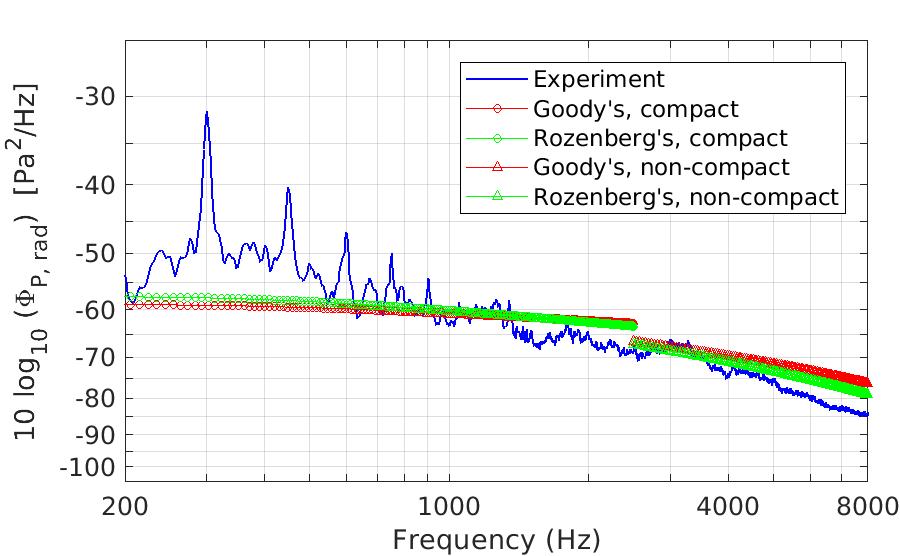}
\caption{Self noise predictions at $\Phi=0.36$ – nine blade fan}
\label{fig:self92}
\end{figure}

\begin{figure}[H]
\centering
\includegraphics[scale=.25]{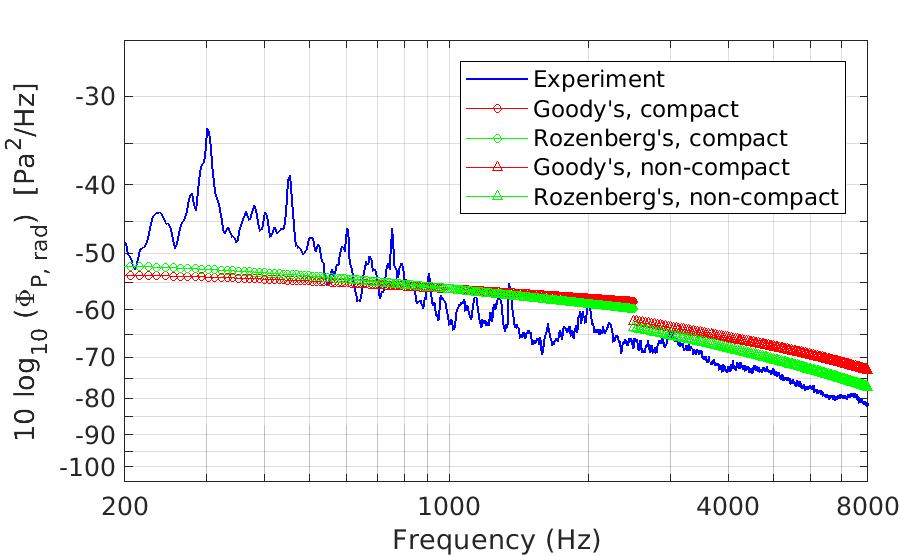}
\caption{Self noise predictions at $\Phi=0.31$ – nine blade fan}
\label{fig:self93}
\end{figure}

\begin{figure}[H]
\centering
\includegraphics[scale=.25]{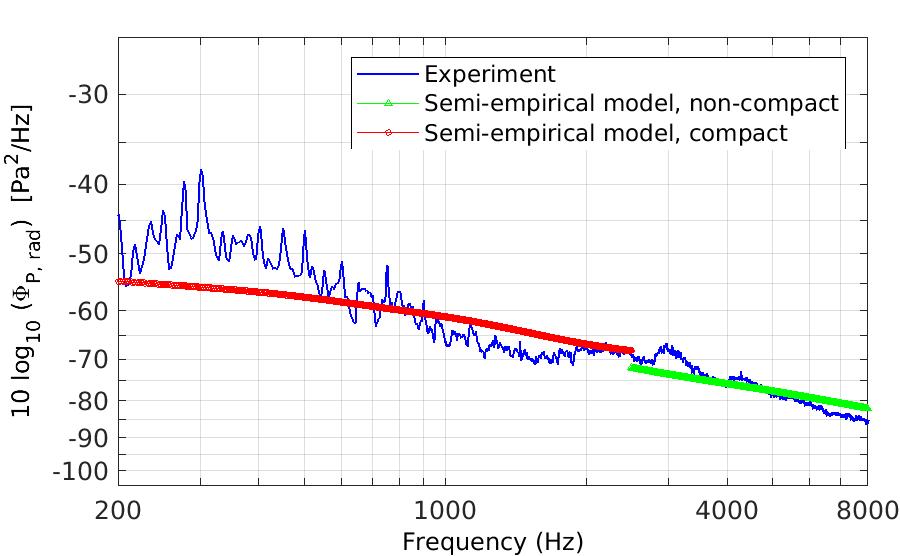}
\caption{Self noise predictions at $\Phi=0.36$ – three blade fan}
\label{fig:self3}
\end{figure}

Acoustic radiation from the three and nine blade fans were measured with the ten microphones as described earlier (Fig. \ref{fig:AFRD}) and post-processed as discussed in the Experimental Setup and Details Section. Experimental results of Figures \ref{fig:self91}-\ref{fig:self3} were taken from reference \cite{davoudi2014aeroacoustic}.  The auto-spectral density of the radiated sound was estimated from the presented model. As it was already explained in the Self-noise Modeling section, the parameters needed for the model were inferred from Figs.~\ref{fig:9cases} and ~\ref{fig:3cases}.




The experimental results and predicted acoustic spectra from the $\Phi=0.49$ case are shown in Fig.\ref{fig:self91}. Note again this is the highest flow rate condition that is considered, where the fan-wake data indicated the flow was attached with relatively thin boundary layers over the airfoils. The experimental spectrum shows both broad-band and tonal characteristics. The tones are at integer multiples of the blade-passing frequency. While these tones are not the focus of the present research, it is notable that they exist in the present, low-Mach number fan with conditions that were intended to be axisymmetric. Similar observations have been made by, for example, Fukano et al \cite{fukano1986effects} and Stephens and Morris \cite{stephens2011measurements}. 

The broadband amplitudes are generally decreasing with frequency over the range noted with a relatively constant slope. The model spectra indicate a relatively good approximation to the overall-amplitude in the mid-higher frequency (i.e., acoustically non-compact) range. The lower frequency range (acoustically compact) shows that the model underestimates the measurements by about 10 dB. It is also noted that the slope of the model very closely approximates the experimental spectra over a relatively wide frequency range (f>1000 Hz). As noted, using either Goody's or Rozenberg's models to estimate the surface pressure spectra do not make a considerable difference in the predicted emissions, and they differ by about 1 dB for $\phi=0.49$



The predicted self-noise for $\Phi=0.36$ shown in Fig. \ref{fig:self92}. The broad-band part of experimentally obtained acoustic spectra has been elevated comparing to $\Phi=0.49$ which was expected. That is, as shown in Fig. \ref{fig:9cases}, the wake for $\Phi=0.36$ is larger than the wake for $\Phi=0.49$, and the broadband noise has also correspondingly increased. The predictions are in better agreement with the experiment, and the maximum discrepancy is about 8 dB between the experimental broadband noise and the predicted self-noise. Also, as noted, the Rozenberg's surface pressure model led to a better estimation for both compact and non-compact parts in both broadband amplitude and it's slope. The Rosenberg's model led to a steeper prediction that agree with experimental data better.

The predicted and experimentally obtained spectra for $\Phi=0.31$ is shown in Fig. \ref{fig:self92}. The broad band noise has increased comparing to the previous cases with higher flow coefficients. Also, it is evident that the tonal noise peaks have almost similar values for all of the cases, but broad-band has increased with flow coefficient decrease and made the peaks look less abrupt. For instance, the peak at 300 Hz has almost a similar magnitude of 33 dB for all three cases studied. The predicted spectrum, especially the ones using the Rozenberg's model, is in very good agreement with the experimental broadband noise spectra. The difference between Goody's and Rozenberg's models is larger for higher frequencies and is at a about maximum of 5 dB. Similar to the previous case, Rozenberg's surface pressure model provides a more accurate estimation for amplitude and slope.

Similar acoustic measurements were made for the three blade fan, and acoustic data for $\Phi=0.36$ was depicted in Fig. \ref{fig:self3}. As expected, the experimentally obtained spectra has smaller values comparing to the nine blade fan, however, a similar descending trend with frequency increase was observed.  The semi-empirical model results (only Goody's surface pressure spectra was used) and the experiment are in very good agreement for both the compact and non-compact parts.

The acoustic semi-empirical model indeed incorporated many approximations and assumptions, however, the results were found to be in reasonable agreement with the experiments for both three and nine blade fans.

\section{Acoustic Scaling}

\begin{figure}[!htb]
\centering
\includegraphics[scale=.32]{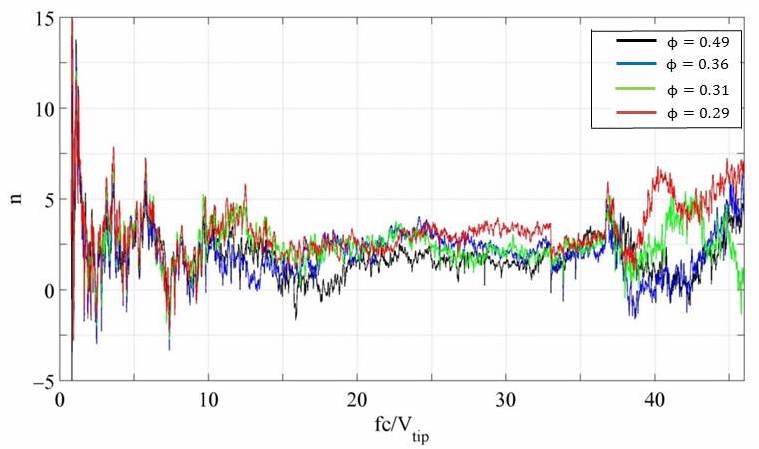}
\caption{Acoustic scaling – nine blade fan}
\label{fig:n}
\end{figure}

Measurements of the emitted fan noise were carried out for different rotational speeds for a given flow and pressure rise coefficients. Using those measurements, the function $"n"$, described in the Introduction section of this paper, was experimentally determined. These results are shown in Fig. \ref{fig:n} for the nine blade fan. Given that the three blade fan self-noise was not sufficiently loud for the lower RPM conditions, the function $"n"$ was not obtained for that fan configuration.

The function $"n"$ was found to vary considerably with frequency. Also, the scaling function values in lower frequencies were quite similar for different operating condition, however, they were different for mid-high frequencies. Very large peaks with a maximum of nearly 15 were observed in low frequencies in Fig. \ref{fig:n}. Similar peaks in the low frequency range were also observed in Stephens and Morris \cite{stephens2011measurements}. The large values of “n” are not well understood. In mid frequencies, the values of n are roughly between 2.5 to 3.5 for different operating conditions which are close to the reported value of 3 by Longhouse \cite{longhouse1976noise}.

\section{Conclusions}

A semi-empirical acoustic model was adopted and modified to predict the emitted self-noise of Rotating Controlled Diffusion Blades (RCDBs) configured as a three and a nine blade axial fan. Four/two operating conditions were selected for the nine/three blade fans, respectively. Wake thickness measurements in a plane near the blades’ TE, and mass flow rate measurements across the fan plane were the inputs to the semi-empirical acoustic model. Surface pressure spectra that was used in the semi-empirical acoustic model was obtained by Goody's and Rozenberg's models separately, and Rozenberg's model offered a more accurate prediction. Constant temperature hot-wire anemometer data were utilized to obtain velocity fluctuations level signatures in the fan blades wake. Those velocity fluctuations level signatures were then used to obtain wake thickness approximations. 
Aeroacoustic measurements were obtained using an array of microphones and the experimental data were compared with the predicted self-noise. The self-noise predictions were found to be in acceptable agreement with the experimental observations. It was also observed that the larger wake thickness values were correlated with stronger acoustic emissions.
Given that aeroacoustics measurements are relatively cumbersome, it could be an efficient procedure to perform single probe hot-wire measurements in the fan wake region and use the presented semi-empirical model to estimate the fan noise. Also, computationally affordable CFD simulations could be employed to estimate the wake characteristics as the inputs to the semi-empirical acoustic model.

\bibliography{bib}
\bibliographystyle{aiaa}

\end{document}